# Type-II Dirac photons at the metasurfaces

Chuandeng Hu[1], Zhenyu Li[2,3], Rui Tong[1], Xiaoxiao Wu[1], Zengzilu Xia[1], Li Wang[1], Shanshan Li[2,3], YingZhou Huang[4], Shuxia Wang[4], Bo Hou[2,3,a)], C. T. Chan[1], Weijia Wen[1,a)]

[1]*Department of Physics, The Hong Kong University of Science and Technology, Clear Water Bay, Kowloon, Hong Kong, China*

[2]*College of Physics, Optoeletronics and Energy & Collaborative Innovation Center of Suzhou Nano Science and Technology, Soochow University, Suzhou 215006, China*

[3]*Key Laboratory of Modern Optical Technologies of Ministry of Education & Key Lab of Advanced Optical Manufacturing Technologies of Jiangsu Province, Suzhou 215006, China*

[4]*Department of Applied Physics, Chongqing University, Chongqing 401331, China*

a) Correspondence and requests for materials should be addressed to Weijia Wen (email: phwen@ust.hk) or Bo Hou (email: houbo@suda.edu.cn).

## Abstract

Topological characteristics of energy bands, such as Dirac/Weyl nodes, have attracted substantial interest in condensed matter systems as well as in classical wave systems. Among these energy bands, the type-II Dirac point is a nodal degeneracy with tilted conical dispersion, leading to a peculiar crossing dispersion in the constant energy plane. Such nodal points have recently been found in electronic materials. The analogous topological feature in photonic systems remains a theoretical curiosity, with experimental realization expected to be challenging. Here, we experimentally realize the type-II Dirac point using a planar metasurface architecture, where the band degeneracy point is protected by the underlying mirror symmetry of the metasurface. Gapless edge modes are found and measured at the boundary between the different domains of the symmetry-broken metasurface. Our work shows that metasurfaces are simple and practical platforms for realizing electromagnetic type-II Dirac points, and their planar structure is a distinct advantage that facilitates applications in two-dimensional topological photonics.

**Introduction**

Topological insulators and semimetals have attracted much interest in current physics research [1-5]. Examples of topological materials include $Z_2$ topological insulators with topological features protected by time-reversal symmetry (TRS), topological crystalline insulators with boundary states protected by crystal symmetries, and Dirac and Weyl semimetals with nodal degeneracy points in momentum space. The conical dispersions associated with nodal points have been classified as either type-I or type-II according to the degree of tilt of the dispersion cones around the degenerate points [6-10]. These features endow the equal-energy-surface at the degenerate point with a single point for type-I nodal points and an intersection of two lines for type-II nodal points.

The topological features in the band structure are not unique to electronic materials but should be realizable for any type of Bloch modes. In fact, non-trivial unique band topologies and the associated protected edge states (ESs), which were first discovered in electronic materials, have been realized in classical dynamical systems such as electromagnetic (EM) [11-27], acoustic [28-32] and mechanical systems [33]. Among all classical wave systems, the topological photonic system has received the most attention due to its foreseeable impact in technology and because the characterization can be carried out using standard equipment. For example, three-dimensional (3D) Dirac points (DPs) and Weyl points have been created in dielectric photonic crystals [15, 19-21]. Recently, the design of type-II DPs using photonic crystals has been theoretically predicted [34-36], but the existence of these DPs has not yet been demonstrated experimentally.

In addition to photonic crystals, electromagnetic metamaterials have been employed to probe the topological features of photonic systems [16-18, 22-25, 31, 37, 38], because the metallic inclusions and the associated resonances afford a greater degree of freedom to the band engineering. In particular, metasurfaces, regarded as two-dimensional (2D) metamaterials, provide an appealing planar design for manipulating an EM wave through controlling the surface resonances and surface waves with technological advantage that is compatible with an on-chip fabrication process [39-43].

In this work, we propose a straightforward way to create 2D type-II DPs and experimentally implement them in a metasurface composed of simple metallic patterns. Using a surface field mapping technique, we experimentally determine the iso-frequency contour (IFC) at the DPs, which is characterized by a linearized touch between two dispersion branches instead of a single point at the degenerate frequency. Furthermore, since the ESs derived from type-II DPs are gapless, ESs have also been observed in momentum space. Our work offers a new strategy to control EM waves, which has the advantage of using a planar architecture and is scalable to terahertz frequencies.

**The emergence of type-II Dirac points in metasurface**

We create type-II DPs at a metasurface by manipulating different bands corresponding to orthogonal electric dipole modes. Our metasurface composed of periodic metallic patterns can be homogenized as an anisotropic dielectric slab under the deep-subwavelength approximation [39-43]. The band emerging from the $\Gamma$ points along the principle axis resembles the lowest guided TE mode of the homogenized anisotropic dielectric slab, in which the electric field is polarized along the in-plane direction perpendicular to the propagation direction. Furthermore, the band corresponding to the guided resonance mode also emerges along the main axis with the electric field polarized parallel to the propagation direction. These two bands are adjustable and can be engineered by the geometry parameters of the metasurfaces, i.e., we can manipulate them individually (see details in Supplementary Notes 1-6). Here, we illustrate such two bands through employing the modified Jerusalem cross as a metallic unit for the metasurface (see several other examples in Supplementary Note 7). The designed geometry details are shown in Fig. 1a, in which the substrate is fabricated using a non-magnetic material with a relative permittivity of 16 and thickness of $d = 2$ mm, with the geometric parameters of the metallic pattern chosen to have a period $D_x = D_y = 2.5$ mm, $a = 2.3$ mm, $b_1 = b_2 = 0.9$ mm, $c = 0.8$ mm, $w_1 = 0.1$ mm, $w_2 = 0.2$ mm, $d = 2$ mm and $t = 18$ $\mu$m. The metallic patterns are printed onto the dielectric slab; a photograph showing the top view of the fabricated sample is shown in the inset of Fig. 1b. We use COMSOL Multiphysics to calculate the fundamental modes of a single unit and the dispersion relations of the first two bands (see Supplementary Note 13). From the electric field (z-component) distributions shown in Fig. 2a, two fundamental modes are purely electric dipole modes with a dipole moment along the x-direction (defined as EDMx in this letter, at 8.52 GHz) and the y-direction (defined as EDMy in this letter, at 16.12 GHz). The two bands intersect at a degenerate point along the $\Gamma$-X direction, as shown in Fig. 2b. The modes located away from the principle axis are not purely EDMx or EDMy, which can be understood by the introduction of a non-diagonal term to the effective permittivity. Furthermore, there is no symmetry-protected degeneracy along a general direction. Therefore, the degeneracy appears at an isolated point in the 3D band diagram, which can be numerically verified, as shown in Fig. 2c. As an important distinction between type-I and type-II DPs, the IFC at the degenerate frequency is a cross shape rather than a single point, as shown in the bottom of Fig. 2c.

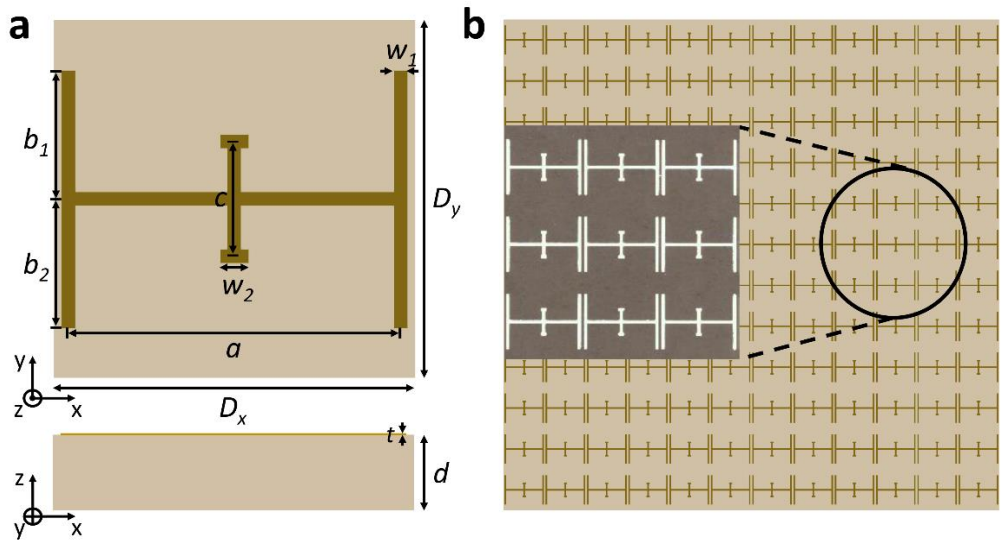

**Figure 1 | Illustration of the metasurface. a** The sketch of a single unit of the designed metasurface with the following parameters: $D_x = D_y = 2.5$ mm, $a = 2.3$ mm, $b_1 = b_2 = 0.9$ mm, $c = 0.8$ mm, $w_1 = 0.1$ mm, $w_2 = 0.2$ mm, $d = 2$ mm, $t = 18$ $\mu$m. **b** The top view of the designed sample with the inset showing a photo of part of the fabricated sample.

## k·p theory

**k·p** theory is adopted to better understand the dispersion around the DPs [23, 35, 44]. The effective Hamiltonian around the type-II DPs can be constructed as follows [45]:

$$H = \boldsymbol{\omega} \cdot \boldsymbol{k} \cdot \hat{I} + v_x k_x \sigma_x + v_y k_y \sigma_y \tag{1}$$

where $\hat{I}$ is the unitary matrix; $\sigma_x$ and $\sigma_y$ are Pauli matrixes; and $v_x$ and $v_y$ represent group velocity along the x- and y-directions, respectively. The parameters $\boldsymbol{\omega}$ ($\omega_x$ and $\omega_y$) denote the tilt of the Dirac cone, in which the tilt parameter is defined as folloes:

$$\hat{\omega} = \sqrt{\left(\frac{\omega_x}{v_x}\right)^2 + \left(\frac{\omega_y}{v_y}\right)^2} \tag{2}$$

in which $\hat{\omega} < 1$ for type-I DPs and $\hat{\omega} > 1$ for the type-II DPs. The dispersion relation can be analytically evaluated from Eq. (1) and reads as follows:

$$f_\pm = \frac{1}{2\pi}\left(\boldsymbol{\omega} \cdot \boldsymbol{k} \pm \sqrt{v_x^2 k_x^2 + v_y^2 k_y^2}\right) \tag{3}$$

here the sign '+ (–)' denotes the upper (lower) band. $k_x = 0$ and $k_y = 0$ marks the location of the degenerate point. The dispersion relation was numerically evaluated (see details in Supplementary Note 11), as shown in Fig. 2d, and the results are in excellent agreement with the numerical calculations.

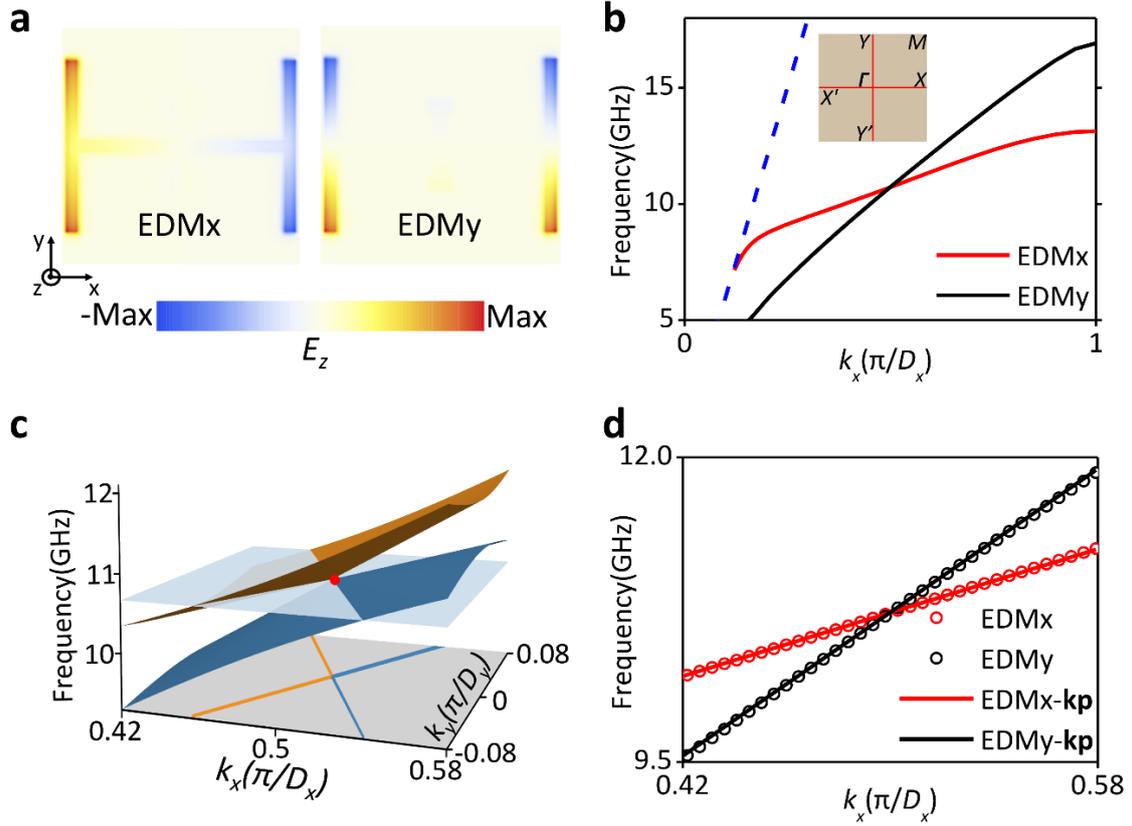

**Figure 2 | Type-II Dirac point at the metasurface. a** Electric field maps of the $E_z$ component of the calculated two modes of the single unit of the designed metasurface. The eigenfrequencies of the electric dipole modes with a dipole moment along the x-direction (EDMx) and y-direction (EDMy) are 8.52 GHz and 16.12 GHz, respectively. **b** The calculated dispersion relations along the $\Gamma$-$X$ direction, with the red and black solid lines denoting EDMx and EDMy, respectively. The blue dashed line is the light line, and the inset shows the first Brillouin zone and high-symmetrical points. **c** 3D view around the DP (red point); the transparent blue surface is the equal-energy surface at the degenerate frequency, with the IFC cut by the surface plotted at the bottom (a cross shape formed by blue and orange lines). **d** Dispersion relation around the DP along the $\Gamma$-$X$ direction obtained via simulations (red and black open circles) and the **k**·**p** method (red and black solid lines).

## Observation of type-II Dirac points

The metasurface consisting of $76 \times 76$ unit cells is fabricated (total size 190 mm $\times$ 190 mm) with geometric parameters as specified in Fig. 1a. The excitation antenna source is set in the center of the edge along the y-direction (see Supplementary Note 8). The dielectric slab is fabricated using a TP1/2 printed circuit board, which is a nonmagnetic material with a relative permittivity of 16 and a tangential loss of 0.001 (at 10 GHz). The metallic patterns are printed on the dielectric slab with 18-μm-thick copper and are covered by a negligible thickness of tin on the copper surface to prevent oxidation. The near-field scanning method is utilized to measure the electric field distribution, and the probing antenna is set above the metasurface at a distance of approximately 1 mm (the measured region is 180 mm $\times$ 180 mm, and the resolution is 2.5 mm $\times$ 2.5 mm). The modes of the first two bands are purely EDMs (EDMx and EDMy); thus, both the $E_x$ and $E_y$ components are measured by adjusting the probing antenna parallel to the x- and y-directions, respectively. We then perform a Fourier transform on the measured real space results, with a total resolution of $72 \times 72$ points in momentum space for each component. After we combine the results for both components in momentum space with a ratio of 1:1, as shown in Fig. 3b, the IFCs of the two bands change with increasing frequency. The IFCs of the first band evolve gradually with the increase in frequency from an elliptical curve to a hyperbola, while the second band emerges from the light cone with IFCs in the form of expanding ellipses. Most importantly, a cross shape between the ellipse and the hyperbola is clearly observed at 10.6 GHz. Although some modes show a small amplitude measured along the $\varGamma$-$Y$ direction, which is due to the finite size of the excitation antenna, and, consequently, inefficient coupling to the $k_x$~0 modes, the experimental results agree well with the numerical results shown in Fig. 3a. To conclude, we experimentally observe the type-II DPs in momentum space using the designed metasurface.

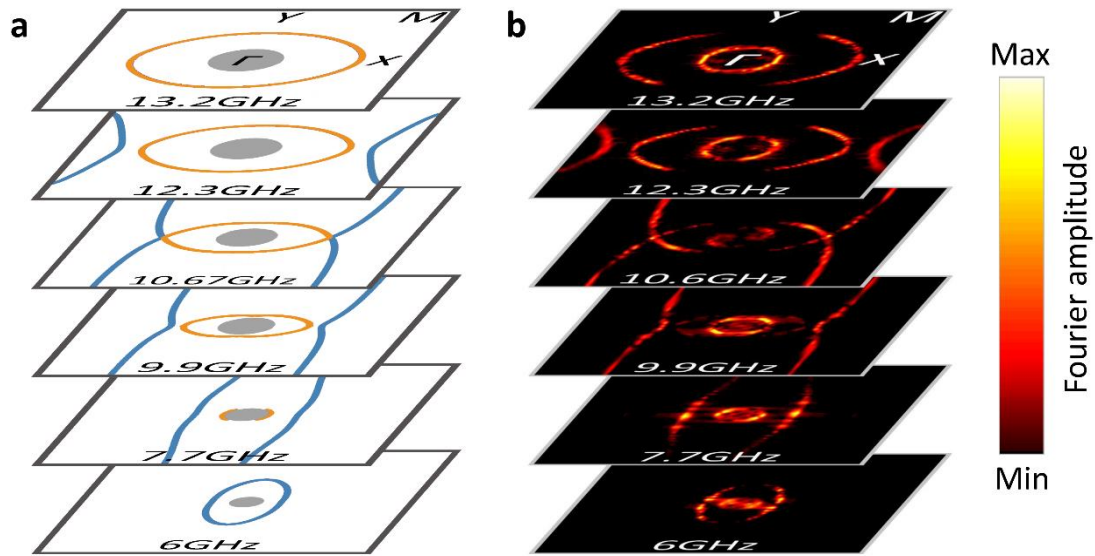

**Figure 3 | Experimental observation of type-II Dirac points.** IFCs in the first Brillouin zone obtained via **a** simulations and **b** experiments. The high-symmetrical points are indicated on the top of the figure. The type-II DPs are seen along the $\varGamma$-$X$ direction at the 10.6 GHz IFC.

## Emergence of edge states

The $Z_2$ topology can be revealed in photonic systems by employing symmetry for both type-I and type-II DPs [35], hence, the ESs corresponding to the emerged DPs can also be created. The DPs in Dirac semimetal are usually protected by a combination of symmetries and topological properties, whereas the type-II DPs are protected, in our case, by the mirror symmetry of the Jerusalem cross pattern about the x-direction. Breaking mirror symmetry (MS) along the x-direction will change the modes from pure EDMx to a combination of EDMx and EDMy, which leads to lifting of degeneracy since the modes in the first two bands are no longer orthogonal. In contrast, the breaking of MS along the y-direction can only induce a shift in the resonant frequency, which does not affect the symmetry of the modes. As such, the type-II Dirac point in Fig. 2b is only protected by the MS along the x-direction. The patterns with broken MS along the x-direction can be divided into two groups by simply exchanging parameters $b_1$ and $b_2$, which gives type A ($b_1 = 0.3$ mm, $b_2 = 1.5$ mm) and type B ($b_1 = 1.5$ mm, $b_2 = 0.3$ mm), as shown in Fig. 4a. The first two bands along $\Gamma$-$X$ ($X'$) are the same for these two patterns which have been evaluated in Fig. 4b, and the degeneracy is lifted. The modes of these two bands located at $k_x = 0.5(\pi/D_x)$ are numerically evaluated with electric field (z-component) distributions shown in Fig. 4c, which shows that the modes are no longer pure EDMx or EDMy (see Supplementary Note 13). The **k p** effective Hamiltonian can be modified as (see details in Supplementary Note 10):

$$H = \boldsymbol{\omega} \cdot \boldsymbol{k} \cdot \hat{\mathrm{I}} + v_x k_x \sigma_x + v_y k_y \sigma_y + \Xi \cdot \sigma_z \quad (4)$$

where $\Xi$ is the shift of the frequency from the mean frequency and $\sigma_z$ is the Pauli matrix. Note that when the symmetry is maintained, the effective Hamiltonian will reduce to Eq. (1) because of the degeneracy ($\Xi = 0$). As for the first band, the Berry curvature can be directly deduced (see details in Supplementary Note 10):

$$\Omega = \pm \frac{v_x v_y \Xi}{2\left(v_x^2 k_x^2 + v_y^2 k_y^2 + \Xi^2\right)^{3/2}} \quad (5)$$

where the sign + and − correspond to $k_x < 0$ and $k_x > 0$, respectively. This condition holds because the MS along the y-direction is retained and will lead to opposite signs for $k_x > 0$ and $k_x < 0$. Here, we numerically evaluate the Berry curvature of the first band of type-A, as shown in Fig. 4d; the valley Chern number can also be evaluated both analytically and numerically, with values of -1/2 for $k_x > 0$ and 1/2 for $k_x < 0$, respectively. The type-B pattern can be regarded as a π rotation of type-A, which leads to the inversion of the Chern number for the same band at the same region because of the nature of the Bloch function. We build a domain wall between these two patterns along the x-direction and categorize them into AB-type and BA-type, corresponding respectively to a type-A pattern placed in the upper domain and lower domain, as shown in Fig. 5a and 5d. The emergence of ESs is analogous to the type-I DPs (valley), which can be understood from the bulk-boundary correspondence [1, 3]. We construct a supercell composed of 20 unit cells (10 type-A patterns and 10 type-B patterns) along the y-direction and set the periodic condition along the x-direction to simulate the projected band diagram along the x-direction (see Supplementary Note 12). The results are shown in Fig. 5b and 5e for AB-type and BA-type stacks, respectively. The regions marked in gray show bulk states, with the modes above the light line along *Γ-X* (*X'*) (dashed blue lines) neglected. The clear emergence of two ESs is indicated by the solid blue lines in Fig. 5b and solid red lines in Fig. 5e. Interestingly, the interface states exist below the lowest bulk states for the AB-type stack, which are denoted by the black solid lines, as shown in Fig. 5b. The interface states are not under topological protection, which can be verified by slightly changing the distance between these two domains (see details in Supplementary Note 9). Moreover, the electric field map for the $E_z$ component and Poynting vector profiles for these two types of ESs are simulated and plotted in Fig. 6, where the opposite energy flux directions corresponding to the opposite wave vectors agree well with the prediction from the Chern numbers. Unlike type-I DPs, the band diagram remains gapless even if the degeneracies are lifted, which results in the emergence of gapless ESs. However, we can observe the ESs in momentum space in analogy to the surface arcs for type-II Weyl points [22].

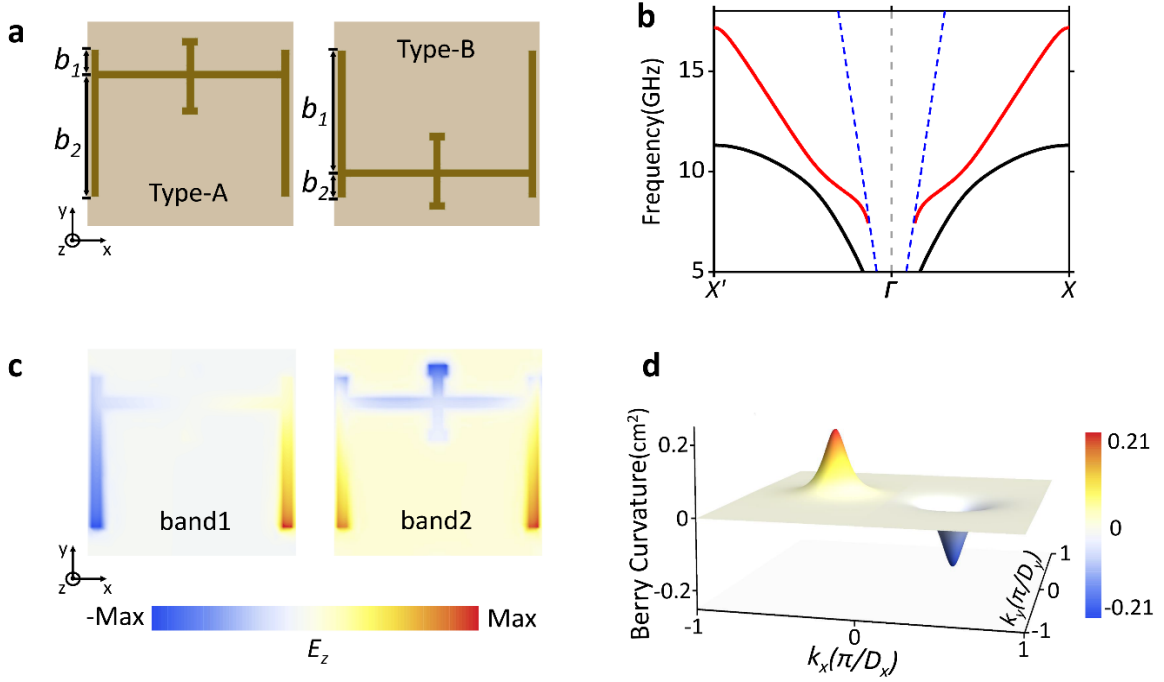

**Figure 4 | Emergence of edge states.** The top view of a single unit of **a** type-A Jerusalem cross with $b_1 = 0.3$ mm, $b_2 = 1.5$ mm and type-B Jerusalem cross with $b_1 = 1.5$ mm, $b_2 = 0.3$ mm, respectively. **b** The dispersion relation of the first two bands along the $\Gamma$-$X$ ($X'$) directions. **c** Electric field maps of the $E_z$ component of the first (band 1) and second (band 2) bands at $k_x = 0.5(\pi/D_x)$. **d** The Berry curvature for the first band of type-A Jerusalem on the first Brillouin zone obtained from the **k**·**p** method.

## Observation of edge states

Each of the two metasurface domains, which consist of 76 ×38 type-A and 76 ×38 type-B patterns (total size 190 mm ×190 mm), are assembled along the x-direction to form AB-type and BA-type stacks, respectively, as illustrated in Fig. 5a and 5d, where the geometric parameters as well as the permittivity of the dielectric slab are the same as in Fig. 4. The excitation source antenna is set at the center of the edge along the y-direction, which is the position of the domain wall (see details in Supplementary Note 8). The near-field scanning is conducted using a measurement region of 180 mm $\times$ 180 mm and resolution of 2.5 mm $\times$ 2.5 mm. Likewise, we perform a Fourier transform to obtain the IFCs for the $E_x$ and $E_y$ components. We project the IFCs along the x-direction by summing over $k_y$ at given $k_x$ and frequency and remove the points located above the light lines along $\Gamma$-$X$ ($X'$) (see Supplementary Note 12). Finally, we normalize the result at each frequency and combine the two components (obtained from the measured $E_x$ and $E_y$ components) at a ratio of 1:1. As shown in Fig. 5c and 5f, two edge states emerge, which agrees well with the numerical results. To conclude, we directly observe the ESs derived from type-II DPs in momentum space even though they are gapless.

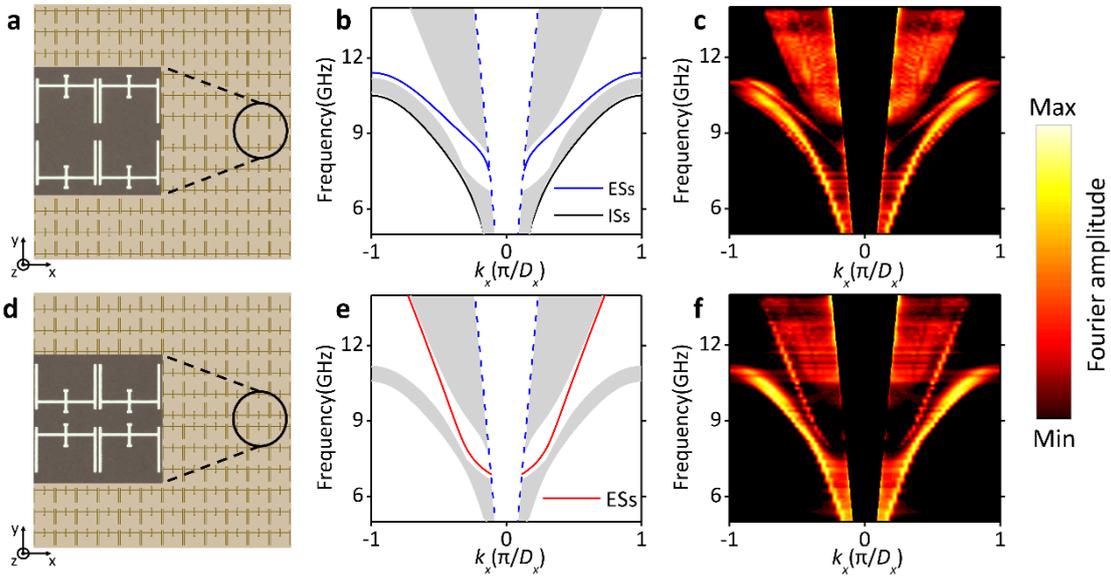

**Figure 5 | Experimental observation of edge states.** The top view of the designed **a** AB-type stack **d** BA-type stack samples; the insets show the corresponding enlarged view of the fabricated sample at the domain walls. **b** and **e** are the project band diagrams along the x-direction for the AB-type and BA-type stacks, respectively. The regions shaded in gray denote

the bulk states, and the blue dashed lines denote the light lines along the $\Gamma$-$X$ ($X'$) directions; the blue solid lines in **b** and the red solid lines in **e** show the edge states (ESs), while the black lines in **b** show the interface states (ISs). **c** and **f** are the projected bands for the AB-type and BA-type stacks obtained from experiment, respectively.

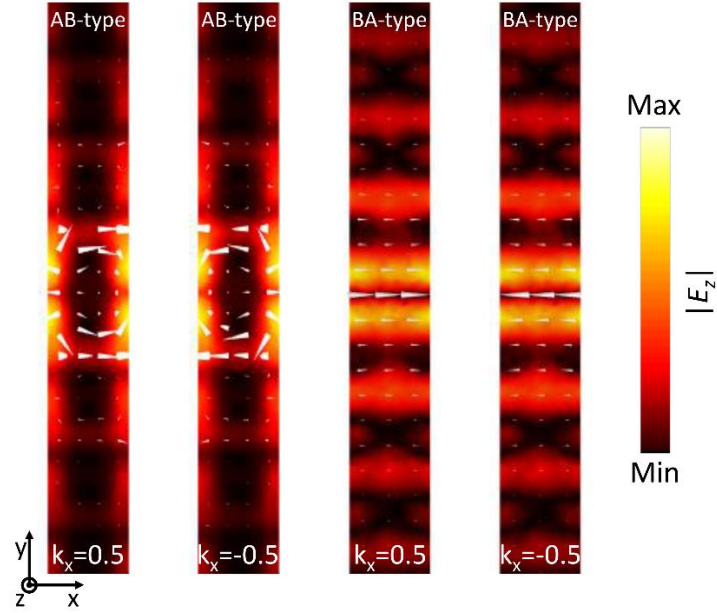

**Figure 6 | Field map of edge states.** Simulated field distribution of the $E_z$ component and Poynting vector profiles for AB-type and BA-type domain walls at opposite wavevectors on the x-y plane at a distance of 1 mm above the surface.

## Discussion

Metasurfaces composed of periodic subwavelength metallic patterns offer superb flexibility in engineering surface modes. Including the EDMs only and mapping the corresponding dispersion relations along the main axis to the specific parameters of the metasurfaces enables the type-II DPs to be successfully created. Our strategy for realizing type-II DPs can, in principle, be scaled to higher frequencies (i.e., terahertz frequencies).

In summary, type-II DPs, as well as derived ESs, are experimentally observed at metasurfaces. Our design provides an ideal platform for probing the topological nature of surface waves and may have potential applications in the design of 2D topological photonic devices.

## Acknowledgements

Chuandeng Hu would like to thank Mingli Chang, Dr. Xiao Xiao and Prof. Jianhua Jiang for the useful discussions. The work is supported by an Areas of Excellence Scheme grant (AOE/P-02/12) from Research Grant Council (RGC) of Hong Kong, the Special Fund for Agro-scientific Research in the Public Interest from Ministry of Agriculture of the Peoples' Republic of China (No. 201303045), and the grants from Natural Science Foundation of China (NSFC) (No. 11474212), and the Priority Academic Program Development (PAPD) of Jiangsu Higher Education Institutions.